\documentclass[a4paper,10pt,twoside]{cpc-hepnp}
\usepackage{CJK,upgreek,fancyhdr}
\usepackage{multicol}
\usepackage{graphicx}
\usepackage{booktabs}
\usepackage{amssymb,bm,mathrsfs,bbm,amscd}
\usepackage[tbtags]{amsmath}
\usepackage{lastpage}

\begin{document}
\begin{CJK*}{GBK}{song}

\fancyhead[c]{\small Chinese Physics C~~~Vol. xx, No. x (201x) xxxxxx}
\fancyfoot[C]{\small 010201-\thepage}

\footnotetext[0]{Received 31 June 2015}

\title{Study of Magnetic Hysteresis Effects in a Storage Ring \\ Using Precision
Tune Measurement\thanks{Supported by National Natural Science Foundation of China (11175180,
11475167) and US DOE (DE-FG02-97ER41033)}}

\author{%
    Wei Li$^{1,2;1)}$\email{wei.li3@duke.edu}
    \quad H. Hao$^{2}$
    \quad Stepan F. Mikhailov$^{2}$
    \quad Wei Xu $^{1}$
    \quad Jing-Yi Li $^{1}$ \\
    \quad Wei-Min Li $^{1}$
    \quad Ying. K. Wu$^{2;2)}$\email{wu@fel.duke.edu}
}
\maketitle
\address{%
$^1$ National Synchrotron Radiation Laboratory, University of Science and Technology
of China, Hefei, 230029, China\\
$^2$ Triangle University Nuclear Laboratory/Physics Department, Duke University, Durham, 27705, USA\\
}
\begin{abstract}
{\normalsize{}With advances in accelerator science and technology
in the recent decades, the accelerator community has focused on the
development of next-generation light sources, for example the diffraction-limited
storage rings (DLSRs), which requires precision control of the electron
beam energy and betatron tunes. This work is aimed at understanding
magnet hysteresis effects on the electron beam energy and lattice
focusing in the circular accelerators, and developing new methods
to gain better control of these effects. In this paper, we will report
our recent experimental study of the magnetic hysteresis effects and
their impacts on the Duke storage ring lattice using the transverse
feedback based precision tune measurement system. The major magnet
hysteresis effects associated with magnet normalization and lattice
ramping are carefully studied to determine an effective procedure
for lattice preparation while maintaining a high degree of reproducibility
of lattice focusing. The local hysteresis effects are also studied
by measuring the betatron tune shifts resulted from adjusting the
setting of a quadrupole. A new technique has been developed to precisely
recover the focusing strength of the quadrupole by returning it to
a proper setting to overcome the local hysteresis effect.}{\normalsize \par}
\end{abstract}

$\text{\,\,}$PACS: 29.20.db, 75.60.Ej

\footnotetext[0]{\hspace*{-3mm}\raisebox{0.3ex}{${\scriptstyle \copyright}$}2013
Chinese Physical Society and the Institute of High Energy Physics
of the Chinese Academy of Sciences and the Institute of Modern Physics
of the Chinese Academy of Sciences and IOP Publishing Ltd}

\begin{multicols}{2}

\section{Introduction}

With advances in accelerator science and technology in the recent
decades, the next-generation synchrotron radiation sources will be
developed with higher brightness, better coherence, and improved beam
control and stability. For versatile, multi-user operation, the accelerator
community has focused on the development next-generation light sources
based upon diffraction-limited storage rings (DLSRs) \cite{Intro_1,Intro_1_2,Intro_1_3,Intro_1_4}.
The development of DLSRs faces a number of scientific and technological
challenges in several areas \cite{Intro_2,Intro_2_2,Intro_2_3,Intro_2_4},
including designing and operating ultra-low emittance lattices (typically,
tens of picometers) with large enough single-particle nonlinear dynamics,
better control of beam parameters and beam instability compared with
the third-generation light sources, good beam lifetime with a limited
dynamic aperture, etc.

Precision control of the electron beam energy and betatron tunes in
the storage ring is critical for the next-generation light sources
to produce photon beams at exact wavelengths as designed, and to realize
good injection efficiency and beam lifetime without losing dynamic
aperture due to lattice focusing errors. The improved beam parameter
control can also greatly benefit the operation of the existing storage
ring based light sources. In the recent years, techniques have been
developed to improve the stability of the electron beam's energy and
orbit by precisely controlling the air temperature in the storage
ring to the level of 0.1 $^{\circ}$C \cite{Intro_3}. In the recent
decade, two advanced techniques have been developed to measure the
absolute energy of the electron beam in the storage ring with a relative
accuracy of few $10^{-5}$: One uses the Resonant Spin Depolarization
(RSD) technique for a high energy electron beam (typically above $1$
GeV) \cite{Intro_4,Intro_5}, and the other uses the Compton scattering
technique for a low energy beam (typically a few hundreds of MeV)
\cite{Intro_6,Intro_7}. However, both techniques are not readily
available to many operational light sources which do not have a means
to accurately measure the electron beam energy during routine operation.

Since 2000s, the field-programmable gate array (FPGA) based digital
bunch-by-bunch longitudinal and transverse feedback systems have been
widely utilized to mitigate beam instabilities at light source storage
rings. The transverse feedback (TFB) has been used to make precision
measurements of betatron tunes for beam studies and for user operation.
For example, at Duke FEL laboratory a precision tune measurement system
was developed based upon the bunch-by-bunch transverse feedback (TFB)
\cite{Intro_8}. With this system, the betatron tunes can be accurately
measured with an rms uncertainty of a few $10^{-5}$, as limited by
the short-term stability of magnet power supplies and the maximum
data memory available in the TFB system. This new diagnostic capability
has made it possible to carry out precision studies of various effects
related to the electron beam energy and betatron tunes.

In this work, we report our recent experimental study of the magnetic
hysteresis effects and their impact on the electron beam energy and
focusing in the storage ring using the TFB based tune measurement
system without the need for absolute beam energy measurements. This
precision tune measurement system has allowed us to advance beam study
in two areas. First, it has enabled us to investigate the magnet hysteresis
effect associated with magnet normalization. The new insight from
this study has allowed us to experimentally determine the effectiveness
of a particular magnet normalization routine, thus providing a way
to devise a more efficient magnet normalization procedure. Second,
using this system to study the tune changes resulted from adjusting
the setting of a quadrupole, we have developed a new technique to
return the quadrupole to a proper setting to precisely recover its
focusing strength, overcoming the local magnetic hysteresis effect.
The main experimental results in these areas will be presented in
the following sections.

\section{Magnetic Hysteresis Effects}

In this Section, we provide a brief review of the ferromagnetic hysteresis.
Magnetic hysteresis describes the nonlinear response of a ferromagnetic
material to the imposed magnetizing field (the $H$-field), producing
hysteresis loops of finite areas when the $H$-field undergoes repetitive
cycles between two fixed values. This lack of reproducibility of the
material magnetization for a given $H$-field, a ``magnetic memory''
effect, is the result of the behaviors of the magnetic domains in
the material, as a certain amount of energy is needed to reorient
the magnetic domains, and/or change the domain wall boundaries and
sizes.

The study of magnetic hysteresis has a long history going back to
late 1800s \cite{Intro_9,Intro_10}. Since mid 1980s however, hysteresis
effects have become a subject of intense research by scientists and
engineers from a variety of research areas, which has significantly
advanced our understanding of the hysteresis phenomenon in general.
Typically, two different approaches are taken. The first one is a
physics based approach which is built upon a certain statistical-mechanical
theory by applying first-order phase transition to analyze various
spin systems \cite{Intro_11,Intro_12,Intro_13}. The second approach
utilizes phenomenological models. Some of the most successful models
are the Preisach model \cite{Intro_14}, Coleman-Hodgdon model \cite{Intro_15,Intro_16},
and Jilies-Atherton model \cite{Intro_17,Intro_19}. The Jiles-Atherton
model developed in early 1980s connects the model parameters directly
with the physical parameters of the magnetic materials, in particular,
the spinning and rotation of the magnetization in these materials
\cite{Intro_19,Intro_20,Erratum}. This model has also been successfully
extended to describe minor hysteresis loops \cite{Intro_22}.

The well-known magnetic hysteresis effect is the DC hysteresis, which
is represented by a quasi-static loop showing the equilibrium position
of the bulk material magnetization. Since 1980s, research has turned
to investigate the nonequlibrium behavior of frequency-dependent hysteresis
in different types of magnetic materials, including the non-conducting
materials such as high frequency ferrites \cite{Intro_23} and electrically
conducting materials \cite{Intro_24}. The frequency-dependency studies
of conducting materials, such as soft iron used to construct accelerator
magnets, takes into account of the eddy current effect. However, it
assumes that the magnetic field penetrates uniformly throughout the
material \cite{Intro_24}.

\begin{center}
\includegraphics[width=4cm]{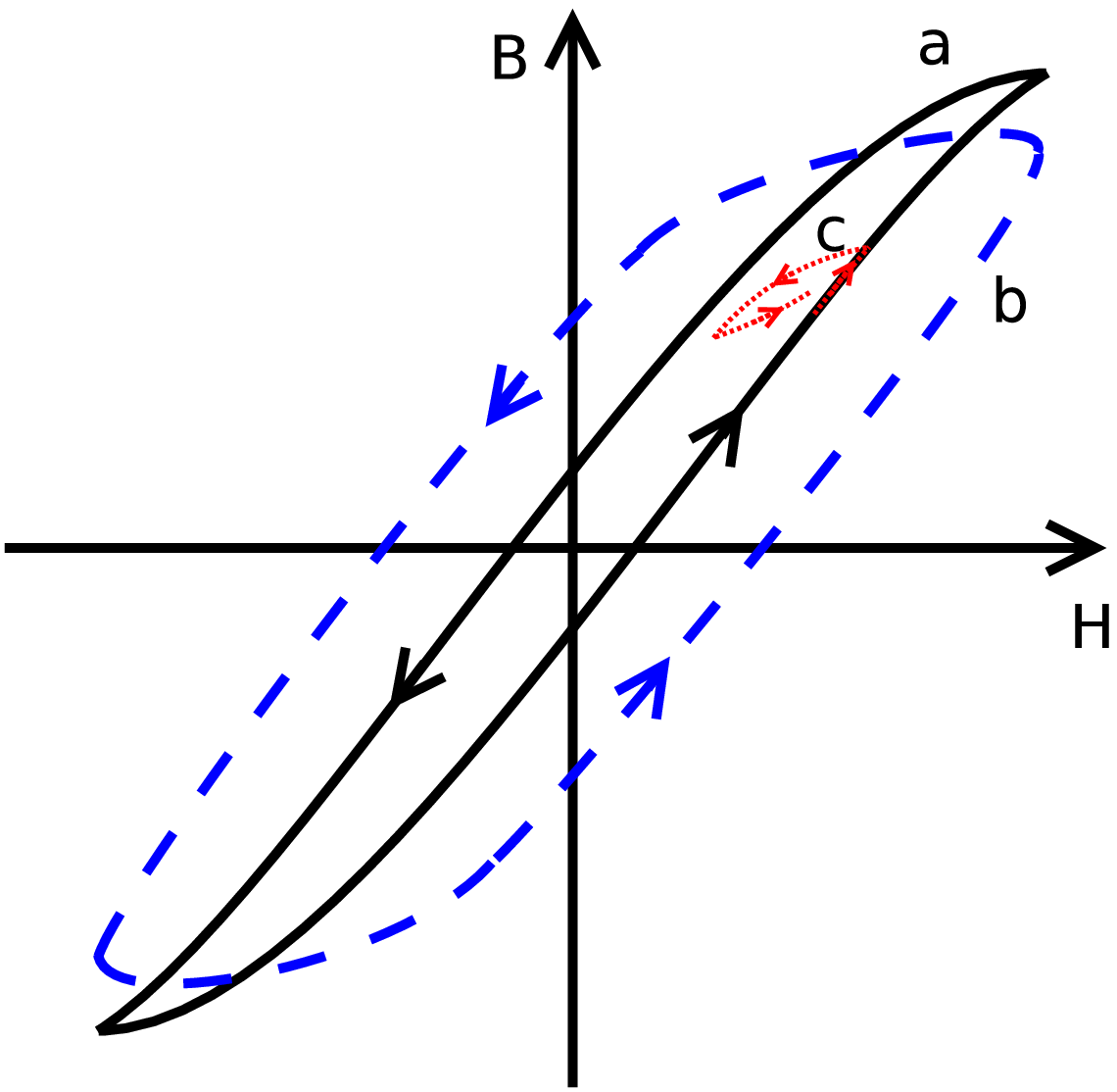}
\figcaption{\label{fig_HysLoop-1} (color online) Hysteresis loops
obtained on a magnetic-curve tracer for soft iron bars. Solid line
a: cycle is performed slowly; Blue dash line b: cycle is performed
fast. Red dot line c: a minor loop. \cite{Intro_11,Intro_25}}
\par\end{center}

Even with significant advances in modeling and understanding magnetic
hysteresis in the recent decades, controlling and managing the adverse
effects associated with magnetic hysteresis remains a very difficult
and challenging task for present and future generation light source
storage rings. These challenges are manifest in several areas, including
(1) precise and accurate control of the accelerator magnet field with
a very small relative uncertainty of $10^{\text{-}4}$ to $10^{\text{-}5}$;
(2) managing magnetic hysteresis associated with the skin effect which
can be important for solid-core magnets even at a low field ramping
rate; (3) dealing with the interference of the magnetic field from
adjacent magnets. In this work, we will report our experimental study
of small changes in magnetic fields related to magnetic hysteresis
using a precision tune measurement system.

\section{Scaling the settings of dipoles and quadrupoles}

A typical way to study magnet hysteresis is to determine the relation
between the $B$-field and $H$-field by measuring the $B\text{-}H$
curve \cite{Chap3-1,Chap3-2}, where the $H$-field is produced by
the coil current and the $B$-field is the nonlinear response to the
$H$-field. However, it is impractical to directly measure the $H$-field
and $B$-field in a well installed facility like Duke storage ring
(DSR). Instead, we use the precision betatron tune measurements to
study the magnet hysteresis. In this section, we provide a brief view
of how dipole and quadrupole magnet field changes can be studied using
the measured tune changes.

The generic Hamiltonian of a charged particle in a normal (non-skew)
quadrupole, a sextupole, or a combined-function quadru-sextupole under
the impulse magnetic field model can be expressed in the Cartesian
coordinate system as \cite{SYLee_APBook,YKWu.Thesis}:

\begin{align}
H(x,y,\delta,p_{x},p_{y},l;s) & \approx\frac{p_{x}^{2}+p_{y}^{2}}{2(1+\delta)}+\frac{1}{2}K_{1}(x^{2}-y^{2})\label{eq2_1-3}\\
+\frac{1}{3}K_{2}(x^{3}-3xy^{2})\nonumber
\end{align}

\noindent where $x,\,y$ are transverse displacements from the designed
orbits; $p_{x\text{,}y}=P_{x,y}/P_{0}$ are the scaled momenta (canonical
to $x,\,y$ respectively) which are given by scaling the momenta $P_{x,y}$
by the designed momentum $P_{0}$; $\delta=\left(P-P_{0}\right)/P_{0}$
is the scaled momentum deviation with $P$ being the particle's momentum,
and $l$ is the path length, $s$ is the independent variable, giving
the particle's orbital position along the storage ring. $K_{1}=\frac{1}{B_{0}\rho_{0}}\frac{\partial B_{y}}{\partial x}\mid_{x=y=0}$,
$K_{2}=\frac{1}{2!}\frac{1}{B_{0}\rho_{0}}\frac{\partial^{2}B_{y}}{\partial x^{2}}\mid_{x=y=0}$
are the strengths of the normal quadrupole and sextupole, respectively,
and $B_{0}\rho_{0}$ is the magnetic rigidity. The related equations
of motion in the transverse directions, for example the $x$-direction,
are

\begin{equation}
\left\{ \begin{aligned}\frac{dx}{ds} & =\frac{p_{x}}{1+\delta}\\
\frac{dp_{x}}{ds} & =-(K_{1}x+K_{2}(x^{2}-y^{2}))
\end{aligned}
\right.\label{eq2_1-4}
\end{equation}
or

\begin{eqnarray}
\frac{d^{2}x}{ds^{2}}+\frac{1}{1+\delta}(K_{1}x+K_{2}(x^{2}-y^{2}))=0.\label{eq2_1-5}
\end{eqnarray}
The linear focusing strength is given by $K_{1}(s)/\left(1+\delta\right)=P_{0}K_{1}(s)/P$.
Therefore, the betatron tune depends on both the strength of the quadrupoles
and the momentum (or energy) of the charged particle. This observations
are still valid even if we take into account the weak focusing provided
by dipole magnets.

The momentum of the charged particle can be determined by the integrated
magnetic field it sees along its closed orbit. Ignoring the betatron
motions (as the average effects is negligible for small amplitude,
fast transverse oscillations), the momentum of the relativistic charged
particle $P$ is given by

\begin{eqnarray}
P=\frac{1}{2\pi c}\intop_{closed\text{\text{-}}orbit}qB_{y}(s)ds\approx\frac{1}{2\pi c}\intop_{dipoles}qB_{y}(s)ds.\label{eq2_1-6}
\end{eqnarray}
Hence the momentum $P$ is proportional to the strength of dipoles.
Consequently, for a storage ring, if the strength of all dipoles and
all quadrupoles are changed by the same relative amount, the betatron
tunes of the beam will remain the same. This is the basic principle
for performing energy ramping in a storage ring or in a booster synchrotron.
This observation means that the betatron tunes will be a good measure
to determine discrepancies in the magnetic field changes between quadrupoles
and dipoles.

The above idea can be tested using a field strength scaling method
by measuring the betatron tune changes as the result of proportionally
changing only the dipole magnetic field (Method 1), or only the quadrupole
magnetic field (Method 2). Method 1 is a commonly used technique to
measure the storage ring natural chromaticity in a conventional storage
ring with only separate-function magnets and with well-corrected beam
orbits in quadrupoles. Changing the magnetic field of all dipoles
alone by the same relative amount will cause a change of the electron
beam energy without altering the orbit in the quadrupoles and other
magnets (such as sextupoles), resulting in a variation of betatron
tunes caused only by the change in effective focusing of all quadrupoles.
In Method 2, the electron beam energy remains unchanged by keeping
the same dipole magnetic field strength. The betatron tunes will vary
as the strength of all other magnets (quadrupoles and sextupoles)
is changed by the same relative amount. In both methods, we expect
that the linear dependency of the betatron tunes on the relative change
in the field strength will give the same value, which corresponds
to the natural chromaticity for a storage ring with separate-function
magnets. In fact, here we are proposing an alternative way to measure
the storage ring natural chromaticity using Method 2.

These effects are measured in the DSR by separately scaling the magnetic
field strength of all dipoles or all quadrupoles/quadru-sextupoles.
As shown in Fig. \ref{fig_scaleBQ}, the betatron tunes show an opposite
response in Method 2 compared to Method 1 by comparing subplot Fig.
2 (a) vs (c) for the horizontal tune ($\nu_{x}$) and Fig. 2 (b) vs
(d) for vertical tune ($\nu_{y}$). The slope of the tune variations
gives the measured natural chromaticity. In the horizontal (vertical)
direction, the tune slopes are -9.75 (-8.54) by
varying dipole fields only, and 9.93 (8.73) by varying quadru-sextupole
fields only. The relative difference of 2\% between two different
methods (in both directions) is remarkably small, recognizing the
fact that the tune stability and accuracy of the tune measurement
system are on the order of few 10$^{\text{-}5}$ (absolute values),
and that the magnetic field stability and controllability is on the
order of few 10$^{\text{-}5}$ (relative). These slope values are
close to the natural chromaticity values from a design lattice which
uses only separate-function magnets $\xi_{x}\thickapprox\text{-}9.8$
and $\xi_{y}\thickapprox\text{-}9.5$. The somewhat larger discrepancy
in the vertical natural chromaticity is likely the result of employing
combined function quadru-sextupoles in the real storage ring which
are also used as weak dipole magnets with a large orbit displacement
in these magnets \cite{Magnet.Design,Quad.Sext.magnet}. In Fig. 2
(e) and (f), the measured tunes are shown as all magnets (both dipoles
and quadrupoles/quadru-sextupoles) are varied by the same relative
amount (up to 0.2\%). The measured tunes remain fairly constant in
this range without showing a particular trend, and the resultant small
tune variations ($\delta\nu_{x}\thickapprox0.0005$, and $\delta\nu_{y}\thickapprox0.0007$,
peak-to-peak) are at a level expected from various experimental limitations
mentioned earlier. This experiment has demonstrated two important
results: first, we have independent and accurate control of electron
beam energy (using dipoles) and focusing strengths (using quadrupole/quadru-sextupoles);
second, the physics idea that scaling the magnetic field in all magnets
can reserve tunes (an idea outlined earlier) works rather well in
this real storage ring with complicated combined-function magnets.

\end{multicols} \ruleup

\begin{center}
\includegraphics[width=16cm]{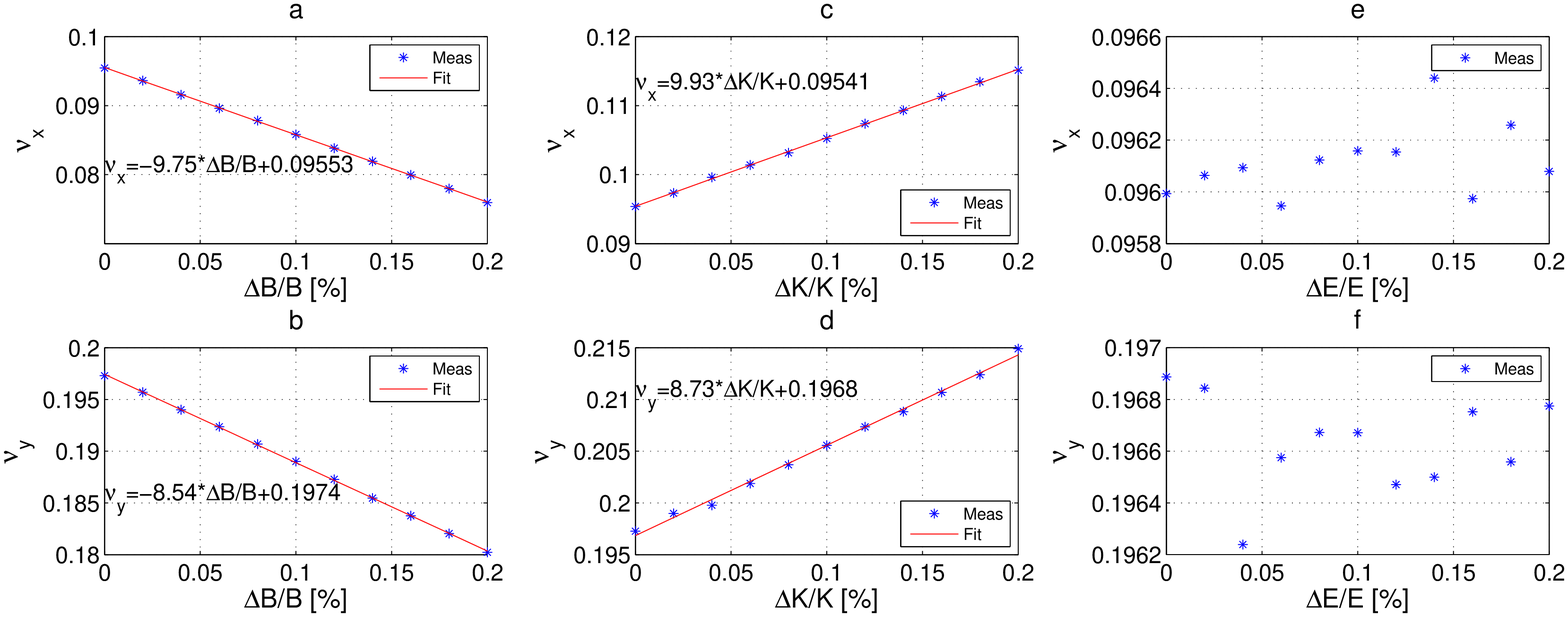}
\figcaption{\label{fig_scaleBQ} (color online) Measured betatron
tune shifts by scaling the strength of the magnets of different types
in the DSR. The measured fractional tunes ($\nu_{x}$ and $\nu_{y}$)
are shown as a function of the change of the relative strength of
the magnetic field: subplots (a) and (b) for all dipole magnets, (c)
and (d) for all quadru-sextupoles, and (e) and (f) for both dipoles
and quadru-sextupoles.}
\par\end{center}

\ruledown \begin{multicols}{2}

\section{Major hysteresis effect}

As mentioned in the previous section, by changing the rate of magnetization
process (commonly known as ``magnet normalization'' in accelerator
operation), a magnet can be magnetized to different levels at the
end of this process. The amount of time needed to properly complete
the magnetization process differs for different types of magnets due
to the differences in their size, shape, material used, and the operation
range of the field. When the magnetization process is made very slowly,
all major magnets (dipoles, quadrupoles, etc.) will have adequate
time to be properly magnetized to reach their expected field values
for operation. However, to save the setup time for operation, in practice,
a faster magnetization process is commonly used. This may result in
some inconsistency between the magnetic fields among dipoles and quadrupoles,
leading to noticeable magnetic lattice variations which will show
up as betatron tune variations.

In a typical storage ring, the lattice preparation is performed to
magnetize all major magnets in two steps. First, the currents in these
magnets are repeatedly ramped between a set of pre-determined minimum
and maximum values for each type of magnets in a so-called magnet
normalization cycle. Second, the magnets are ramped toward the final
set-points for beam operation (termed \textquotedblleft lattice ramping\textquotedblright{}
in this work). Using the DSR, we have investigated these two processes
separately using the precision tune measurement system. In this section,
we describe our experimental procedures and report our main findings.

For the DSR, all main magnets, including dipoles and quadrupole/quadru-sextupole,
were carefully characterized by performing magnetic measurements on
each individual magnet. These measurements were used to determine
the mappings between the measured magnetic fields and magnet coil
currents; these mappings were later implemented in an advanced physics
based real-time accelerator controls system \cite{Physics.Based.Control.}.
For the magnet measurements, the magnet field ramping was usually
performed slowly in order to obtain a stable quasi-static magnetization
curve. Ideally for acceleration operation, the magnets should be prepared
using a ramp rate which is the same as or close to that used in the
measurements when possible. However, for practical reasons (e.g. to
save time and/or to ramp all types of magnets simultaneously), the
magnet normalization for operation is typically carried out at a different
rate from that used in the measurements for many storage rings. In
the DSR, the ramping rate of a typical normalization cycle is $r_{n}=r{}_{0}=10\text{ MeV/s}$,
and that of lattice ramping is four times slower $r_{l}=r_{0}/4=2.5\text{ MeV/s}$.
Both are faster than the typical rate used for magnet measurements
$r_{meas}\thickapprox2\text{ {{MeV/s}}}$.

\end{multicols} \ruleup

\noindent \begin{center}
\includegraphics[clip,width=16cm]{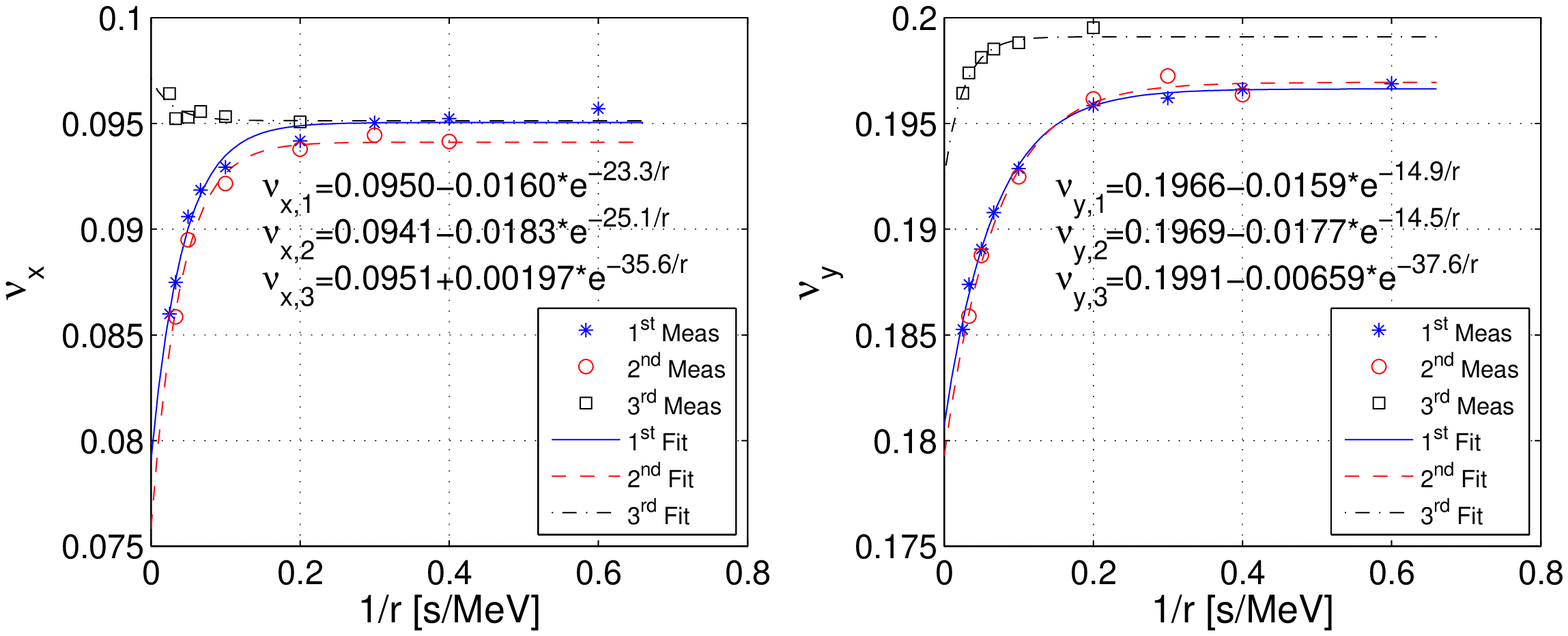}
\figcaption{\label{fig_NormRamp} (color online) Measured fractional
betatron tunes of a 638 MeV DSR lattice prepared with different ramping
rates used for the lattice setup. The first experiment: $r_{n,1}=r{}_{l,1}$
and varied together; the second experiment: $r_{n,2}=r_{0}$ and $r{}_{l,2}$
is varied; the third experiment: $r{}_{l,3}=r_{0}/4$ and $r_{n,3}$
is varied. Three sets of experiments were carried out in three days.}
\par\end{center}

\ruledown \begin{multicols}{2}

To find out the impact of the magnet ramping rates used in the normalization
cycle and lattice ramping on the final betatron tunes, three experiments
are performed on the DSR by varying the magnet ramping rate in each
process. In the first experiment, the rates of normalization cycle
and lattice ramping are the same $r_{n,1}=r{}_{l,1}$ for lattice
preparation, and the betatron tunes are measured for the lattices
prepared with the ramping rate varied from $r{}_{0}/6$ to $4r{}_{0}$.
In the second experiment, the rate of the normalization cycles is
kept constant $r{}_{n,2}=r_{0}$ for all the measurements, while the
lattice ramping rate is changed in a range. In the third one, the
rate of lattice ramping is fixed, but the rate of each normalization
cycle is varied. The ramping rates for these experiments are summarized
in Table \ref{tab_ExpSetting}.

To increase the lattice consistency in these experiments, the storage
ring lattice is prepared with a proper procedure in which magnets
are ramped to 638 MeV settings after three normalization cycles. To
reduce the impact of beam current on the measured tunes, the beam
current in the storage ring is controlled between 4 -- 4.5 mA and
measured tunes are properly corrected to take into account the transverse
impedance related tune shift with beam current \cite{ChaoA.Beam.Decay}.

\noindent \begin{center}
\tabcaption{ \label{tab_ExpSetting} Different ramping rates for
magnet normalization ($r_{n}$) and lattice ramping ($r_{l}$) are
used in the three experiments. All ramping rates are expressed in
terms of $r_{0}=10\text{ MeV/s}$.}%
\begin{tabular}{|c|c|c|}
\hline
Experiment & $r_{n}$ (normalization) & $r_{l}$ (lattice ramp)\tabularnewline
\hline
\hline
$1^{st}$ Exp & $r_{n,1}$: $r{}_{0}/6\text{ to }4r_{0}$ & $r_{l,1}=r$$_{n,1}$\tabularnewline
\hline
$2^{nd}$ Exp & $r_{n,2}=r_{0}$ & $r_{l,2}{{\text{: }r_{0}/4\text{ to }3r_{0}}}$\tabularnewline
\hline
$3^{rd}$ Exp & $r_{n,3}{{\text{: }r_{0}/2\text{ to }4r_{0}}}$ & $r_{l,3}=r_{0}/4$\tabularnewline
\hline
\end{tabular}
\par\end{center}

In the first experiment, the magnet ramping rates for normalization
and the last step lattice ramping are kept the same, and changed from
$1.7$ ($r_{0}/6$) to $40$ ($4r_{0}$) MeV/s. The measured betatron
tunes are shown as starred data points in Fig. \ref{fig_NormRamp},
with each data point representing a complete process of lattice preparation,
beam injection, and steady state of storage ring operation for measurements.
When the ramping rate is slow enough, roughly $r\le5$ MeV/s, the
measured betatron tunes (both horizontal and vertical) have very consistent
values; the measured tunes are found to approach asymptotic values
as the ramping rate $r$ approaches zero. This tune behavior can be
modeled using an exponential decay as a function of $1/r$, $\nu(r)=\nu_{0}+A\,e^{-B/r}$,
where $\nu_{0}$ is the asymptotic value, $A$ and $B$ are two characteristic
parameters describing the hysteresis effects (see the inset formulas
for $\nu_{x,1}$ and $\nu_{y,1}$ in Fig. \ref{fig_NormRamp}). The
experimental finding of the asymptotic tune values confirms the existence
of the stationary hysteresis curve for magnets, which can be approached
using a proper normalization procedure at a slow ramping rate.

When the ramping rate is increased from $5$ MeV/s to $40$ MeV/s,
the measured betatron tunes decrease significantly. With faster ramping,
the magnet field strength is expected to be slightly smaller than
the stationary value, which can cause a tune increase due to weaker
dipole magnets (see Section 3) and a tune decrease due to weaker quadrupole
magnets. While both effects are present in our experiment, from the
observed betatron tune reduction in both directions at a higher ramping
rate, we can conclude unambiguously that the quadrupole effect dominates---the
loss of the relative quadrupole strength is greater than the loss
of the relative dipole strength with fast ramping. A total of about
eight hours was used to carry out the first experiment, which involved
the study of nine $638$ MeV lattices prepared using different ramping
rates.

Faster magnet normalization is used in routine operation to save the
lattice preparation time. The second experiment is designed to validate
this time-saving procedure developed based upon operational experience.
In this experiment, the ramping rate for the normalization cycles
is fixed at $10$ MeV/s ($r_{n,2}=r_{0}$) the same as in routine
operation, while the last step lattice ramping rate $r_{l,2}$ is
varied from $2.5$ MeV ($r_{0}/4$ ) to $30$ MeV/s ($3r_{0}$). The
measured betatron tunes (circled data points in Fig. 3) as a function
of the ramping rate ($r_{l,2}$) show a very similar trend as in the
first experiment, which is easily seen by comparing two sets of fitting
curves, $\nu_{x,1}(r)$ vs $\nu_{x,2}(r)$, and $\nu_{y,1}(r)$ vs
$\nu_{y,2}(r)$. Again, when $r_{l,2}\le5$ MeV/s, the measured tune
values are reasonably close to their respective asymptotic values,
and at a faster rate, a significant tune decrease is observed. Using
the fitting model $\nu=\nu_{0}+A\,e^{-B/r}$, we can make an estimate
for the reasonable ramping rate. To achieve a certain tune repeatability
$\delta\nu$, the maximum ramping rate is given by $r_{\max}=B/\ln(\left|A\right|/\delta\nu)$.
In the horizontal direction, a reasonable choice for tune repeatability
is $\delta\nu_{x}=1\times10^{-4}$, which leads to $r_{1,{\rm max}}=4.6$
MeV/s (exp. \#1) and $r_{2,{\rm max}}=4.8$ MeV/s (exp. \#2). In the
vertical direction, a reasonable choice for tune repeatability is
$\delta\nu_{y}=2\times10^{-4}$ (the overall focusing in the vertical
direction is about one half of that in the horizontal, with $Q_{x}/Q_{y}\thickapprox0.46$,
where $Q_{x,y}$ are the total betatron tunes), which leads to $r_{1,{\rm max}}=3.4$
MeV/s (exp. \#1) and $r_{2,{\rm max}}=3.2$ MeV/s (exp. \#2). These
estimated rates confirm that the routine lattice preparation procedure
used in operation ($10$ MeV/s for normalization and $2.5$ MeV/s
for lattice ramping) is a conservative way to achieve the desirable
tune repeatability. In fact, a more general observation can be made:
with reasonably fast normalization (in this example, $r_{n}=10$ MeV/s),
the tune reproducibility is mainly determined by the rate of final
lattice ramping.

To explore the possibility of even faster normalization, a third experiment
is designed to change the ramping rate for normalization ($r_{n,3}$)
while keeping the same slow lattice ramping ($r_{l,3}=2.5$ MeV/s).
The measured tunes are shown as squared data points in Fig. \ref{fig_NormRamp}.
We notice a shift of the vertical tune asymptotic (by about $0.002$)
which is likely caused by changes of the storage ring operational
environment as this experiment was conducted on a different day, including
temperature changes, less consistent orbits in the combined function
quadru-sextupoles, etc. These changes did not impact the overall trend
of the measured tunes in the same day, which always showed a high
level of consistency (see Fig. \ref{fig_NormRamp}). Comparing the
fitting curves among the three experiments, we notice that in the
third experiment, the prefactor coefficient $A$ is much smaller (by
a factor of ${2.4}$ to $9.3$) and the coefficient in
the exponent $B$ is larger (by a factor of $1.4$ to $2.6$). Furthermore,
using the same tune reproducibility criteria, larger maximum ramp
rates are found with $r_{max}=12$ MeV/s from $\nu_{x,3}(r)$ and
$11$ MeV/s from $\nu_{y,3}(r)$. In the horizontal direction, when
the normalization rate is larger than about $12$ MeV/s, the tune
is shifted toward higher as the rate increases, a very surprising
finding which indicates the dipole hysteresis effect is now dominating
over the quadrupole effect.  In addition to reproducing desired lattice
tunes, the lattice preparation is also aimed at producing an electron
beam with accurate energy, which requires the hysteresis effect of
the dipole magnets to be minimized, hence it is important not to carry
out the normalization cycle too rapidly.

\section{Local hysteresis effect}

In routine operation or during machine study, the setting of a dipole
or a quadrupole/quadru-sextupole is sometimes adjusted by a small
amount for compensation or correction. After bringing back the magnet
setting, the magnet field seen by the electron beam does not return
to the previous value due to local hysteresis, and the discrepancy
in the field introduces perturbation to linear and non-linear beam
dynamics. In this section, we will present our observations of this
type of local hysteresis effect in adjusting quadrupole magnets using
a tune based technique, as well as a compensation scheme for this
local hysteresis effect which is very important for characterizing
the lattice.

\subsection*{Beta-function measurement}

To characterize a storage ring lattice, a series of direct beta-function
measurements can be carried out. A widely used method to directly
measure beta-functions of an individually powered quadrupole is to
change its focusing strength by a small amount and measure the corresponding
betatron tunes. The beta-function averaged for the length of this
quadrupole can be expressed as

\begin{equation}
\beta=\frac{2}{\Delta k_{1}L_{eff}}\frac{\cos\left(2\pi\,\nu_{0}\right)-\cos\left(2\pi\,\nu'\right)}{\sin\left(2\pi\,\nu_{0}\right)}\label{eq:5.1-1-beta.function.original}
\end{equation}
where $\Delta k_{1}$ is the quadrupole strength variation, $L{}_{eff}$
is the effective length of quadrupole, $\nu{}_{0}$ and $\nu'$ are
the fractional betatron tunes of the unperturbed and perturbed lattices,
respectively.

Like most accelerators, the dipoles and quadrupoles in DSR were fully
characterized during the magnetic measurements before installation
along the up-curve of hysteresis loop. The physics based control system
at DSR can properly set the magnets (dipoles and quadrupoles) along
the direction of increasing strength \cite{Physics.Based.Control.}.
For example, one can always increase the strength of a quadrupole
by changing its setting by $\Delta k_{1}$, positive for focusing
quadrupole and negative for defocusing quadrupole. After a beta-function
measurement, the setting of the quadrupole magnet is usually brought
back to the original value to return the lattice back to the original
state. However, the focusing strength seen by the electron beams has
changed slightly due to local hysteresis in this quadrupole, leading
to betatron tune shifts.

\noindent \end{multicols} \ruleup

\begin{center}
{\footnotesize{}\includegraphics[width=16cm]{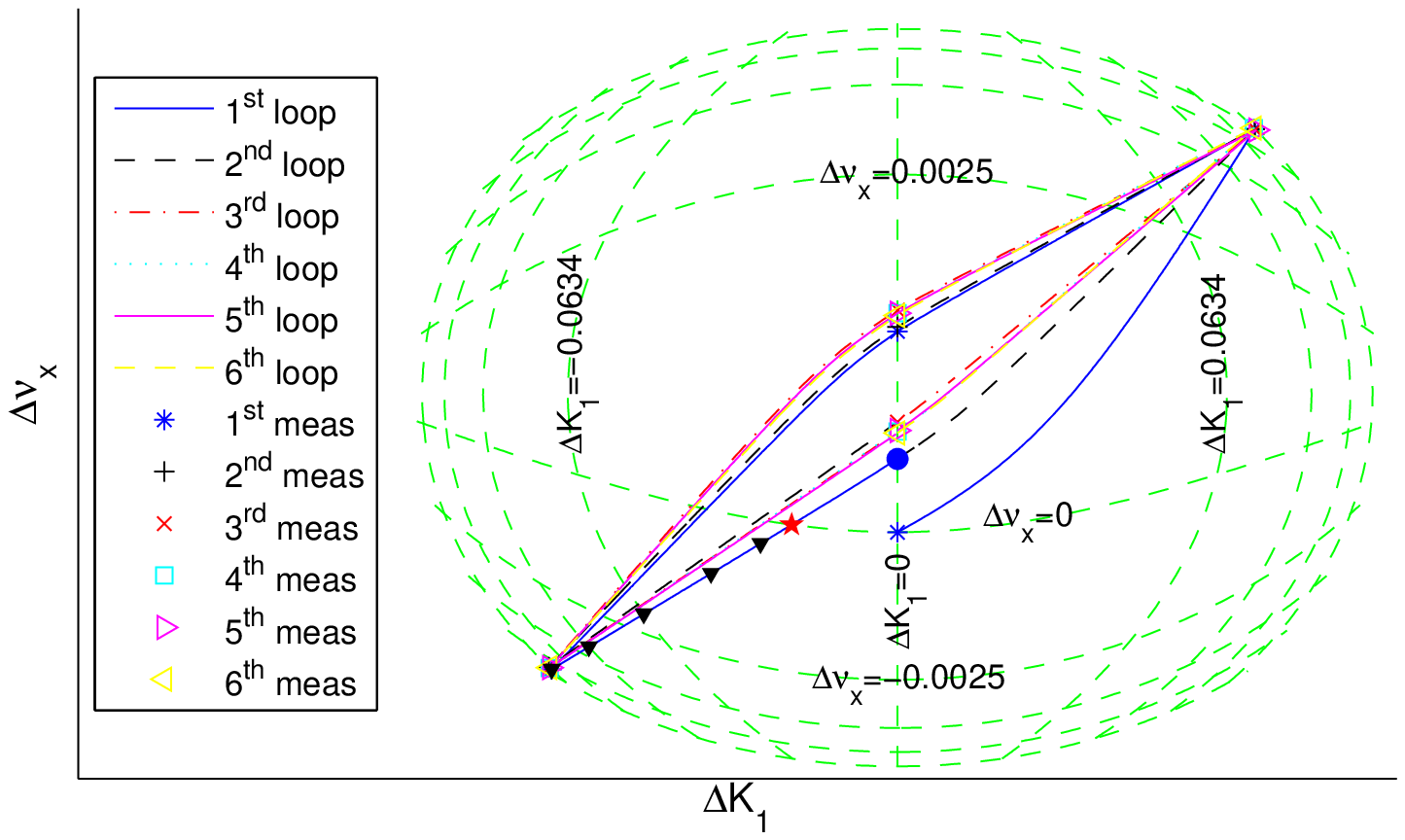}}
\figcaption{\label{fig_minorloopfisheye} (color online) Measure
horizontal betatron tune variations as the setting of quadrupole E04QF
is adjusted in small local loops. The storage ring is operating at
$638\text{ MeV.}$ The grid lines are equally spaced with a tune separation
$\Delta\nu=0.0025$ and $K_{1}$ separation $\Delta K_{1}=0.0634$
between any two adjacent grids. }
\par\end{center}

\noindent \ruledown \begin{multicols}{2}

An example measurement is taken using focusing quadrupole E04QF in
the DSR. In this measurement, the E04QF setting is changed in a local
loop of $\Delta\vec{K}{}_{1}=\left[0,\,\Delta k_{1},\,0,\,\text{--}\Delta k_{1},\,0\right]$,
where $\Delta k_{1}$ is small compared to the nominal quadrupole
setting of $K{}_{1}$. The horizontal tune is measured in sequence
for each quadrupole setting, and this local loop is repeated 6 times.
The beam current dependent tune shift is taken into account so that
all measured tune values are properly adjusted for a fixed beam current
\cite{ChaoA.Beam.Decay}. To visualize the small shifts along the
local loops, a fish-eye plot technique has been developed. The horizontal
tune variation with the change of focusing strength setting is shown
in Fig. \ref{fig_minorloopfisheye}, and the related data are collected
in Tab. \ref{tab_minortune}. Given the resolution of the TFB based
tune measurement system of about $4\times10^{-5}$, the non-reproducibility
of betatron tunes after completing a local loop can be easily measured.

\begin{center}
\tabcaption{ \label{tab_minortune} Measured horizontal betatron
tune variations $\Delta\nu_{x}$ by ramping quadrupole E04QF in local
loops. }{\footnotesize{}}%
\begin{tabular*}{80mm}{@{\extracolsep{\fill}}c@{\extracolsep{\fill}}cccc}
\toprule
Loop  & \multicolumn{4}{c}{Tune variations $\Delta\nu_{x}$ {[}$10^{-3}${]}}\tabularnewline
\midrule
{$\Delta K_{1}${[}$\frac{1}{m^{2}}${]}}  & 0  & 0.254  & 0  & -0.254 \tabularnewline
\midrule
1  & 0 & 9.66  & 1.03 & -8.83\tabularnewline
2  & 0.48 & 9.66  & 1.06 & -8.67\tabularnewline
3  & 0.65 & 9.62 & 1.15 & -8.70\tabularnewline
4  & 0.61 & 9.71 & 1.13 & -8.73\tabularnewline
5  & 0.61 & 9.59 & 1.13 & -8.72\tabularnewline
6  & 0.60 & 9.71 & 1.12 & -8.72\tabularnewline
\bottomrule
\end{tabular*}
\par\end{center}{\footnotesize \par}

Eq. \ref{eq:5.1-1-beta.function.original} is employed to calculate
the average horizontal beta-function ${ {{\beta}}}_{{{{x.i}}}}$
using the two consecutive betatron tunes in each segment of the local
loops in sequence, i.e., using the $i$-th and $\left(i+1\right)$th
measured betatron tune. The calculated beta-functions are compared
with the first measurement $\beta_{x,1}$ (calculated using the $1^{st}$
and the $2^{nd}$ tune measurements), which is measured along in the
main hysteresis curve and considered to be a ``true'' beta-function
value of the quadrupole, and the relative differences are shown in
Fig. \ref{fig_relativebeta}. As shown in the case of $i=2$, if this
quadrupole setting is changed along the reversed direction in the
beta-function measurement, the measured beta-function is found to
be more than $15\%$ smaller than the true value. However, if the
measurement is executed a second time along the up-curve of the hysteresis
loop without normalizing this quadrupole magnet (case of $i=5$),
an error of about 5\% is introduced. Therefore, to measure the beta-function
correctly, only betatron tunes measured along the up-curve of the
main hysteresis curve should be used. It is also observed in Fig.
\ref{fig_minorloopfisheye} that the local loops approach a quasi-static
limit after repeating the loops a few times, leading to a repetitive
pattern in terms of relative beta-function differences shown in Fig.
\ref{fig_relativebeta}. That means a quasi-static minor hysteresis
curve can be developed by repeating a specified magnetization cycle.
This observation is consistent with practices used to normalize major
magnets --- the main storage ring magnets are repeatedly ramped between
a set of maximum and minimum settings without having to reach the
``full saturation'' and ``negative full saturation'' (a procedure
not practical for accelerators).

\noindent \begin{center}
{\footnotesize{}\includegraphics[width=8cm]{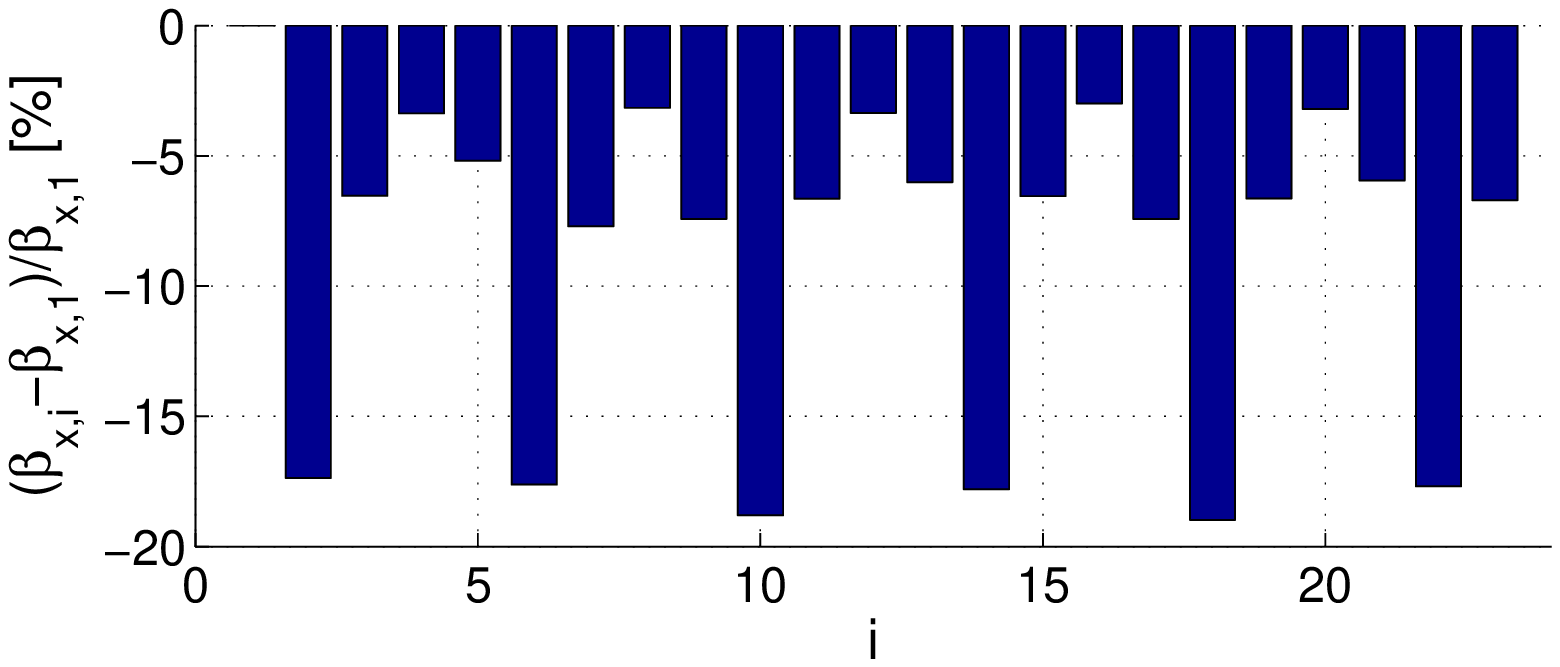}
}\figcaption{\label{fig_relativebeta}Relative differences between
the first measured beta-function $\beta_{x,1}$ along the up hysteresis
curve and additional beta-functions obtained using the tune changes
in subsequent segments of the loops. Each loop is separated into four
segments, for example, $\beta_{x,2}=2.06$ m is calculated using the
second and third betatron tunes measured in the first local loop;
$\beta_{x,4}=2.41$ m is calculated using the forth betatron tune
measured in the first local loop and the first betatron tune measured
in the second loop. }
\par\end{center}

\subsection*{Local hysteresis compensation}

After performing a beta-function measurement in a quadrupole, the
setting of the magnet is returned to the original value by completing
a closed local loop. However, the small discrepancy of the magnetic
field in the quadrupole leads to a small beta beating in the storage
ring, which can be estimated using

\begin{equation}
\frac{\Delta\beta\left(s\right)}{\beta\left(s\right)}=-\frac{2\pi\Delta\nu}{\sin\left(2\pi\,\nu_{0}\right)}\cos\left(2\psi\left(s\right)-2\psi\left(s_{0}\right)-2\pi\,\nu_{0}\right)\label{eq:5.2-1-tune.shift.on.beta}
\end{equation}
where $s_{0}$ is the location of the quadrupole, $s$ is an arbitrary
location in the storage ring, $\Delta\nu=\nu'-\nu_{0}$ is the betatron
tune shift, and $\Delta\psi=\psi\left(s\right)-\psi\left(s_{0}\right)$
is the phase advance between $s$ and $s{}_{0}$. Even though both
the tune shift and beta beating caused by one single measurement are
small, these effects will accumulate over a series of beta-function
measurements. The accumulated effects can become significant to make
accurate beta-function measurements impossible. Thus, a local hysteresis
compensation scheme is needed.

To precisely restore the storage ring lattice, the quadrupole magnetic
field needs to be precisely recovered, which can be verified by checking
the return of the betatron tunes. In our experiment with quadrupole
E04QF, its final focusing should be set to $\Delta K_{1}=\Delta k_{1,f}$
(the red star in Fig. \ref{fig_minorloopfisheye}) to recover the
betatron tune rather than to return the quadrupole setting $\Delta K_{1}=0$
(the blue circle in Fig. \ref{fig_minorloopfisheye}).

The final quadrupole setting $\Delta k_{1,f}$ can be approached by
adjusting the quadrupole strength step by step along the last (or
fourth) segment of the first local loop. A typical process to determine
$\Delta k_{1,f}$ for a quadrupole in the DSR is as follows: After
ramping the quadrupole along a local loop of $\Delta\vec{K}_{1}=[0,\,\Delta k_{1},\,0,\,-\Delta k_{1}]$
and measuring corresponding betatron tunes $[\nu_{0},\,\nu_{1},\,\nu_{2},\,\nu_{3}]$,
the quadrupole strength is first set to $\Delta K_{1}={-}\Delta k_{1}/2$
and the corresponding betatron tune $\nu_{4}$ is measured. By comparing
the initial betatron tune $\nu_{0}$ with $\nu_{4}$, the next adjustment
of the quadrupole strength $\delta k$ can be estimated using the
tune difference $\Delta\nu=\nu_{0}-\nu_{4}$ and the slope $S$ between
two previous measurements ($\nu_{3}$ and $\nu_{4}$) with $\delta k=\Delta\nu/S$.
To keep the quadrupole settings along the up hysteresis curve, the
final adjustment of the setting is done in multiple steps with decreased
step sizes. With a few iterations, the value of $\Delta k_{1,f}$
is determined when the tune value is within a small range from the
original value $\nu_{0}$ (typically $\delta\nu=4\times10^{-5}$ or
terminated after four iterations), as illustrated by the black solid
triangles in Fig. \ref{fig_minorloopfisheye}. Thus, the special local
loop of $\Delta\vec{K}_{1}=[0,\,\Delta k_{1},\,0,\,-\Delta k_{1},\,\Delta k_{1,f}]$
is found for a quadrupole to recover the lattice after a beta-function
measurement. Due to the local hysteresis, the value of $\Delta k_{1,f}$
for a quadrupole depends on the nominal setting $K_{1}$ and the variation
step $\Delta k_{1}$ used for the beta-function measurement.

\noindent \end{multicols} \ruleup

\noindent \begin{center}
{\footnotesize{}\includegraphics[width=16cm]{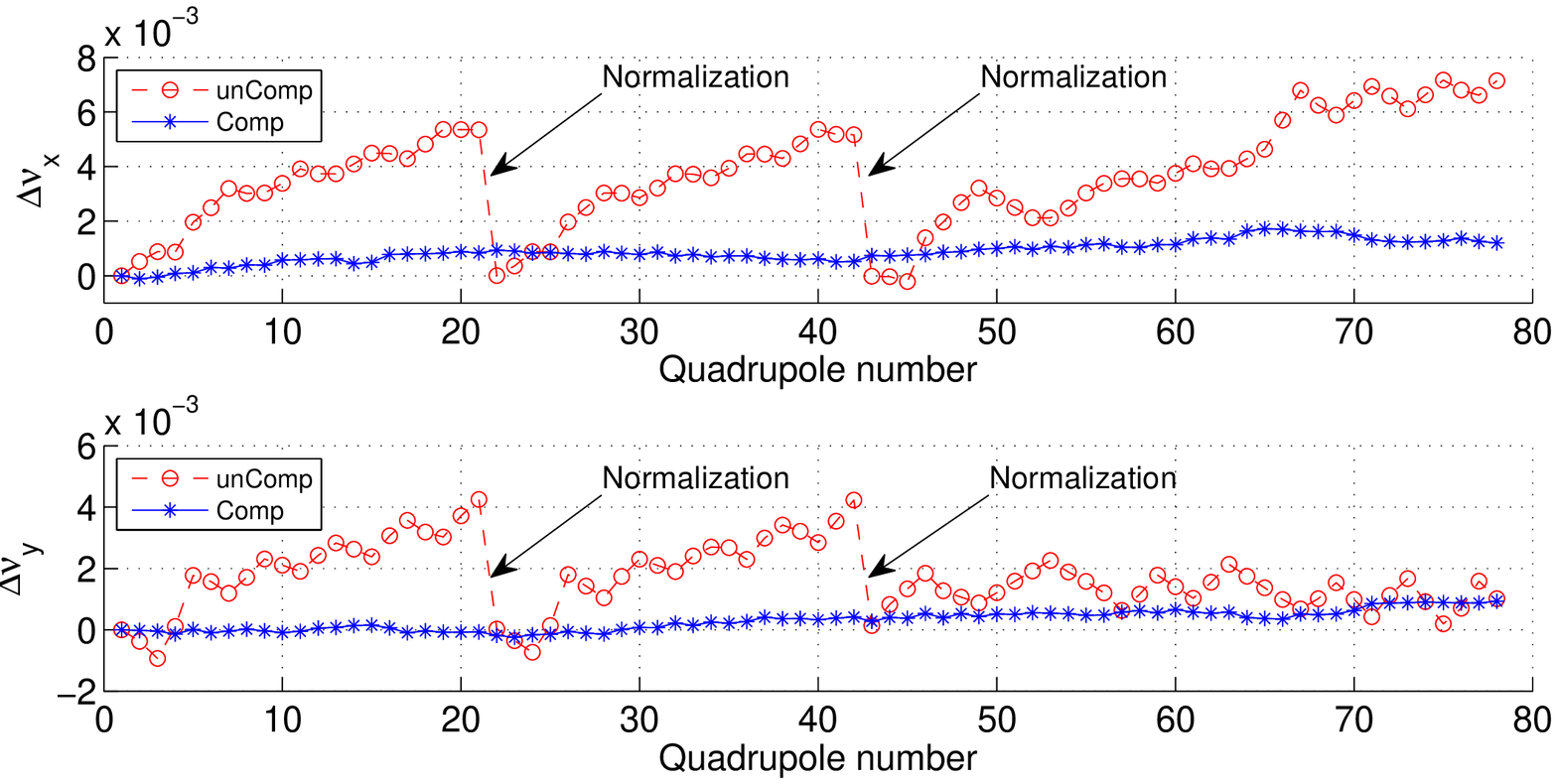}
}\figcaption{\label{fig:5.2-1-accumulated.tune.shift} (color online)
Accumulated tune shifts in the lattice characterization for the DSR
with a total of 78 quadrupoles. The red circles are tune shifts before
performing local hysteresis compensation; two normalization cycle
are used to keep he tune shifts in an acceptable range. Blue stars
are the accumulated tune shifts with the local hysteresis compensation
scheme. The storage ring is operated at $638\text{ MeV.}$}
\par\end{center}

\noindent \ruledown \begin{multicols}{2}

This local hysteresis compensation scheme is developed and utilized
in the lattice characterization for the DSR, which employs direct
beta-function measurements of all quadrupoles. This scheme works rather
well in reducing the accumulated tune shifts, as shown in Fig. \ref{fig:5.2-1-accumulated.tune.shift}.
In the earlier measurements, the DSR needed to be normalized twice
to keep the accumulated tune shifts within $7\times10^{-3}$ in $\nu_{x}$
and $5\times10^{-3}$ in $\nu_{y}$ in the beta-function measurements
for all 78 quadrupoles. But with the implementation of the local hysteresis
compensation, the tune shifts are reduced to $2\times10^{-3}$ in
$\nu_{x}$ and $1\times10^{-3}$ in $\nu_{y}$ without having to normalize
the storage ring.

\section{Summary and discussion}

In the circular accelerators, discrepancies in magnetic fields due
to the hysteresis effect in ramping magnets introduce changes to the
electron beam energy as well as the storage ring magnetic lattice.
Even though the hysteresis effect can be reduced by using proper magnet
materials and the lamination technique, it cannot be eliminated completely.
This can be a problem for the next generation light source like DLSRs,
which requires better control of the electron beam parameters and
linear/non-linear dynamics of the electrons. This work is aimed at
better understanding the magnet hysteresis effects in the storage
ring and, furthermore, developing new methods to precisely control
the electron beam in operation and machine study.

The main impact of magnet hysteresis has been studied experimentally
using the TFB based precision tune measurement system. The first part
of this research is carried out to demonstrate independent control
of the electron beam energy in the DSR by changing the relative strength
of all dipole magnets, as well as independent control of lattice focusing
by varying the relative strength of all quadrupole magnets (quadru-sextupoles).
Both are useful techniques to measure the natural chromaticity of
the storage ring. In the second part of this research, the hysteresis
effect associated with the normalization cycles and lattice ramping
in the process of lattice preparation has been studied in detail using
a series of measurements with three different ways to vary the magnet
ramping rates. Our experimental results show that the rate for the
normalization cycles can be increased to a certain level to save the
lattice preparation time, and good betatron tune reproducibility can
be realized with relatively slow final lattice ramping toward the
operation set-point. Based upon the experimental results, we have
provided a way to estimate the maximum ramping rates for both normalization
cycles and lattice ramping.

To understand the local hysteresis effect, we have studied the tune
shifts due to focusing strength discrepancy resulted from quadrupole
adjustments in the beta-function measurement. Observation in this
experiment suggests that the beta-function should be measured along
the up-curve of the main hysteresis loop. It also indicates that a
quasi-state local hysteresis loop can be approached by repeating a
special local magnetization routine in the magnet. A local hysteresis
compensation scheme has been carefully developed to closely recover
lattice focusing by bringing back the betatron tunes after a beta-function
measurement. With the application of this scheme in the lattice characterization
for the DSR, the accumulated betatron tune shifts are significantly
reduced without having to normalize the storage ring magnets during
the process of measuring the beta-functions for all 78 quadrupoles.
\\

\acknowledgments{We would like to thank the engineering and technical staff at
DFELL/TUNL for their support of this research work. This work was
supported by }\textit{National Natural Science Foundation of China
(No. 11175180, 11475167)}\emph{ and DOE Grant (No. DE-FG02-97ER41033).
One of the authors (Wei Li) also would like to thank the China Scholarship
Council (CSC) for supporting his research visit at Duke University.}

\end{multicols}

\vspace{-1mm}
\centerline{\rule{80mm}{0.1pt}}
\vspace{2mm}
\begin{multicols}{2}

\end{multicols}

\clearpage{}

\end{CJK*}
\end{document}